\documentclass[11pt]{article}

\usepackage{amsmath}
\usepackage{amsfonts}
\usepackage{amssymb}
\usepackage{times}
\usepackage{latexsym}
\usepackage{epsfig}
\usepackage{amsthm}
\usepackage{natbib}
\usepackage{hyperref}
\usepackage{color}

\begin{document}

\title{Least Squares estimation of two \\ ordered monotone regression curves}

\author{Fadoua Balabdaoui$^{(1,2)}$, Kaspar Rufibach$^{(3)}$ and Filippo Santambrogio$^{(1)}$ \\
$^1$ CEREMADE \\
Universit\'e de Paris-Dauphine \\
Place du Mar\'echal de Lattre de Tassigny \\
75775 Paris CEDEX 16, France \\*[0.2cm]
$^2$ Universit\"at G\"ottingen \\
Institut f\"ur Mathematische Stochastik \\
Goldschmidtstrasse 7 \\
37077 G\"ottingen \\*[0.2cm]
$^3$
Universit\"at Z\"urich \\
Institut f\"ur Sozial- und Pr\"aventivmedizin \\
Abteilung Biostatistik \\
Hirschengraben 84 \\
8001 Z\"urich \\*[0.2cm]}

\maketitle

\begin{abstract}
In this paper, we consider the problem of finding the Least Squares estimators of two isotonic
regression curves $g^\circ_1$ and $g^\circ_2$ under the additional constraint that they are ordered; e.g.,
$g^\circ_1 \le g^\circ_2$.  Given two sets of $n$ data points $y_1, \ldots, y_n$ and $z_1, \ldots,z_n$ observed
at (the same) design points, the estimates of the true curves are obtained by minimizing the weighted Least
Squares criterion $L_2(a, b) = \sum_{j=1}^n (y_j - a_j)^2 w_{1,j}+ \sum_{j=1}^n (z_j - b_j)^2 w_{2,j}$ over the
class of pairs of vectors $(a, b) \in \mathbb{R}^n \times \mathbb{R}^n $  such that $a_1 \le a_2 \le \ldots\le a_n
$,  $b_1 \le b_2 \le \ldots\le b_n $,  and $a_i \le b_i, i=1, \ldots,n$. The characterization of the estimators is
established. To compute these estimators, we use an iterative projected subgradient algorithm, where the
projection is performed with a ``generalized'' pool-adjacent-violaters algorithm (PAVA), a byproduct of this
work. Then, we apply the estimation method to real data from mechanical engineering.\bigskip
\end{abstract}


\newtheorem{theorem}{Theorem}[section]
\newtheorem{lemma}[theorem]{Lemma}
\newtheorem{corollary}[theorem]{Corollary}
\newtheorem{proposition}[theorem]{Proposition}
\newtheorem{example}[theorem]{Example}
\newtheorem{definition}[theorem]{Definition}

\newenvironment{Theorem}{\begin{theorem}\sl}{\end{theorem}}
\newenvironment{Lemma}{\begin{lemma}\sl}{\end{lemma}}
\newenvironment{Corollary}{\begin{corollary}\sl}{\end{corollary}}
\newenvironment{Proposition}{\begin{proposition}\sl}{\end{proposition}}
\newenvironment{Example}{\begin{example}\rm}{\end{example}}
\newenvironment{Definition}{\begin{definition}\rm}{\end{definition}}

\let\proglang=\textsf
\newcommand{\pkg}[1]{{\fontseries{b}\selectfont #1}}
\let\code=\texttt

\newcommand{\mycite}[1]{{\small \sc \citeNP{#1}}}

\setlength{\marginparsep}{0mm}
\newcommand{\kr}[1]{{\noindent \bf \color{red}#1}}       
\newcommand{\red}[1]{{\noindent \color{red}#1}}
\newcommand{\blue}[1]{{\noindent \color{blue}#1}}

\newcommand{\blanco}[1]{ }

\def\bs{\boldsymbol}
\newcommand{\ED}{{\mathbb F}_n}                     
\newcommand{\Xin}{X_1,\ldots,X_n}                   
\newcommand{\mean}{\frac{1}{n} \sum_{i=1}^n}        %
\newcommand{\sumn}{\sum_{i=1}^n}                    %
\def\d{\, \mathrm d}                                
\newcommand{\norm}[1]{\left\Vert #1 \right\Vert}    
\newcommand{\SupN}[2]{\Vert #1 \Vert_\infty^{#2}}   
\def\Var{\mathop{\rm Var}\nolimits}
\def\Bias{\mathop{\rm Bias}\nolimits}
\def\Mse{\mathop{\rm Mse}\nolimits}
\def\roc{\mathop{\rm ROC}\nolimits}
\def\auc{\mathop{\rm AUC}\nolimits}

\def\est{{\hat f}_n}                                
\def\lest{{\hat \varphi}_n}                         
\def\dest{{\hat F}_n}                               
\def\hest{{\hat \lambda}_n}                         
\newcommand{\Hol}[3]{\HH^{#1,#2}(#3)}               
\newcommand{\knots}[1]{\mathop{\rm knots(#1)}\nolimits}

\def\toD{\to_{d}}
\def\eqD{\stackrel{d}{=}}
\def\toP{\to_{\rm p}}
\def\toas{\stackrel{\mathrm{as}}{\to}}

\newcommand{\ve}[1]{\bs{#1}}
\newcommand{\mat}[1]{\ve{#1}}       

\def\iid{i.i.d.~}
\newcommand{\ie}{i.e.~}
\newcommand{\eg}{e.g.~}
\newcommand{\wrt}{w.r.t.~}
\newcommand{\as}{a.s.~}

\newcommand{\ruck}[1]{\strut\hspace{#1cm}}

\def\A{{\bf A}}
\def\B{{\bf B}}
\def\C{{\bf C}}
\def\D{{\bf D}}
\def\E{{\bf E}}
\def\F{{\bf F}}
\def\G{{\bf G}}
\def\H{{\bf H}}
\def\J{{\bf J}}
\def\K{{\bf K}}
\def\M{{\bf M}}
\def\O{{\bf O}}
\def\P{{\bf P}}
\def\S{{\bf S}}
\def\T{{\bf T}}
\def\U{{\bf U}}
\def\V{{\bf V}}
\def\W{{\bf W}}
\def\x{{\bf x}}
\def\X{{\bf X}}
\def\y{{\bf y}}
\def\Y{{\bf Y}}
\def\z{{\bf z}}

\def\AA{{\cal A}}
\def\BB{{\cal B}}
\def\CC{{\cal C}}
\def\DD{{\cal D}}
\def\EE{{\cal E}}
\def\FF{{\cal F}}
\def\GG{{\cal G}}
\def\HH{{\cal H}}
\def\II{{\cal I}}
\def\JJ{{\cal J}}
\def\KK{{\cal K}}
\def\LL{{\cal L}}
\def\MM{{\cal M}}
\def\NN{{\cal N}}
\def\OO{{\cal O}}
\def\PP{{\cal P}}
\def\QQ{{\cal Q}}
\def\RR{{\cal R}}
\def\SS{{\cal S}}
\def\TT{{\cal T}}
\def\UU{{\cal U}}
\def\VV{{\cal V}}
\def\WW{{\cal W}}
\def\XX{{\cal X}}
\def\YY{{\cal Y}}
\def\ZZ{{\cal Z}}

\def\alp{\alpha}
\def\gam{\gamma}
\def\Gam{\Gamma}
\def\del{\delta}
\def\Del{\Delta}
\def\eps{\varepsilon}
\def\kap{\kappa}
\def\lam{\lambda}
\def\Lam{\Lambda}
\def\sig{\sigma}
\def\Sig{\Sigma}
\def\th{\theta}
\def\Th{\Theta}
\def\om{\omega}
\def\Om{\Omega}

\def\bea{\begin{eqnarray*}}
\def\eea{\end{eqnarray*}}
\def\be{\begin{equation}}
\def\ee{\end{equation}}
\def\bean{\begin{eqnarray}}
\def\eean{\end{eqnarray}}
\def\barr{\begin{array}}
\def\earr{\end{array}}
\def\bdes{\begin{description}}
\def\edes{\end{description}}
\def\bi{\begin{itemize}}
\def\ei{\end{itemize}}

\def\nin{\noindent}
\def\nn{\nonumber}
\def\npb{\nopagebreak}

\def\comp{^{\rm c}}
\def\trans{^\top}
\def\cd{ \, \vert \, }
\def\Bcd{ \, \Big\vert \, }

\def\Bl{\Bigl}
\def\Br{\Bigr}
\def\la{\leftarrow}
\def\op{o_{\rm p}}
\def\Op{O_{\rm p}}
\def\da{\downarrow}
\def\ua{\uparrow}
\def\by{\times}
\def\til{\widetilde}
\def\hat{\widehat}
\def\setm{\setminus}
\def\subs{\subset}
\def\sups{\supset}
\def\lg{\langle}
\def\rg{\rangle}

\def\Ex{\mathop{\rm I\!E}\nolimits}
\def\Pr{\mathop{\rm I\!P}\nolimits}

\def\N{\mathbb{N}}
\def\Q{\mathbb{Q}}
\def\R{\mathbb{R}}
\def\Z{\mathbb{Z}}

\newcommand{\dom}{\mathrm{dom}}
\def\const{\mathop{\rm const}}
\def\arctanh{\mathop{\rm arctanh}}
\def\artanh{\mathop{\rm artanh}}
\def\argmax{\mathop{\rm arg\,max}}
\def\argmin{\mathop{\rm arg\,min}}
\def\ave{\mathop{\rm ave}}
\def\Beta{{\rm Beta}}
\def\Bin{{\rm Bin}}
\def\Borel{{\rm Borel}}
\def\cl{{\rm cl}}
\def\conv{\mathop{\rm conv}\nolimits}
\def\cone{\mathop{\rm cone}\nolimits}
\def\Cov{\mathop{\rm Cov}\nolimits}
\def\Corr{\mathop{\rm Corr}\nolimits}
\def\det{\mathop{\rm det}}
\def\diag{\mathop{\rm diag}\nolimits}
\def\diam{\mathop{\rm diam}\nolimits}
\def\dist{\mathop{\rm dist}\nolimits}
\def\dim{\mathop{\rm dim}\nolimits}
\def\Exp{\mathop{\rm Exp}\nolimits}
\def\extr{\mathop{\rm extr}\nolimits}
\def\Geom{\mathop{\rm Geom}\nolimits}
\def\Hyp{{\rm Hyp}}
\def\id{\mathop{\rm id}\nolimits}
\def\Leb{\mathop{\rm Leb}\nolimits}
\def\lin{{\rm lin}}
\def\Med{\mathop{\rm Med}\nolimits}
\def\Median{{\rm Median}}
\def\Poiss{{\rm Poiss}}
\def\rang{{\rm rang}}
\def\rank{{\rm rank}}
\def\sign{\mathop{\rm sign}}
\def\spann{{\rm span}}
\def\Std{\mathop{\rm Std}\nolimits}
\def\supp{{\rm supp}}
\def\trace{\mathop{\rm trace}}
\def\TV{\mathop{\rm TV}\nolimits}
\def\Var{\mathop{\rm Var}\nolimits}
\def\Mean{\mathop{\rm Mean}\nolimits}

\def\et{\quad\mbox{and}\quad}
\def\und{\quad\mbox{und}\quad}
\def\oder{\quad\mbox{oder}\quad}
\def\fuer{\quad\mbox{f"ur }}
\def\fueralle{\quad\mbox{f"ur alle }}
\def\fuerein{\quad\mbox{f"ur ein }}
\def\mit{\quad\mbox{mit }}
\def\falls{\quad\mbox{falls }}

\def\ed{\end{document}}

\newcommand{\sg}{${\bf Subgradient}$}
\newcommand{\bam}{${\bf BoundedAntiMean}$}
\newcommand{\ap}{Averaging Property }

\newcommand{\pr}[1]{{\it \noindent Proof#1.}}
\newcommand{\re}{{\it \noindent Remark. }}

\section{Introduction and motivation} \label{intro}
Estimating a monotone regression curve is one of the most classical estimation problems under shape
restrictions, see e.g. \cite{brunk_58}. A regression curve is said to be isotonic if it is monotone
nondecreasing. We chose in this paper to look at the class of
isotonic regression functions. The simple transformation $ g \to- g$ suffices for the results of this paper to carry over to the antitonic class.

Given $n$ fixed points $x_1, \ldots, x_n$, assume that we observe $y_i$ at $x_i$ for $i=1, \ldots, n$. When the
points $(x_i, y_i)$ are joined, the shape of the obtained graph can hint at the increasing monotonicity of the
true regression curve, $g^\circ$ say, assuming the  model $y_i = g^\circ(x_i) + \varepsilon_i$, with
$\varepsilon_i$ the unobserved errors. This shape restriction can also be a feature of the scientific problem at
hand, and hence the need for estimating the true curve in the class of antitonic functions. We refer to
\cite{4b_72} and \cite{robertson_88} for examples. The weighted Least Squares estimate of $g^\circ$ in the class
of isotonic functions taking $y_i$ at $x_i$ is the unique minimizer of the criterion
\bean
L(a) &=& \sum_{i=1}^{n} w_i (y_i - a_i)^2 \label{L}
\eean
over
the class of vectors $a \in \mathbb{R}^n$ such that $a_1 \le a_2 \ldots \le a_n$ where $w_1 > 0,w_2 > 0, \ldots, w_n > 0$
are given positive weights. In what follows, we will say that a vector $v \in \mathbb{R}^n$ is increasing or isotonic if $v_1 \le \ldots  \le v_n$, and use the notation $v \le w$ for $v, w \in \mathbb{R}^n$ if the inequality holds componentwise.

It is well known that the solution $a^*$ of the Least Squares problem in \eqref{L} is given by the so-called min-max formula; i.e.,
\bean
a^*_i = \max_{s \le i} \min_{t \ge i} Av(\{s,\ldots, t\}) \label{eq: minmax1}
\eean
where  $Av(\{s,\ldots, t\}) = \sum_{i =s}^t y_i w_i / \sum_{i=s}^t w_i$ (see e.g. \citealp{4b_72}).

\cite{eeden_57a, eeden_57b} has generalized this problem to incorporate known bounds on the
regression function to estimate; i.e., she considered minimization of $L$ under the constraint
\bean
a_L \le a \le a_U, \label{BoundedRegab}
\eean
for two increasing vectors $a_L$ and $a_U$. As in the classical setting, the solution of this problem admits
also a min-max representation. The PAVA can be generalized to efficiently compute this solution
and has been implemented in the \texttt{R} package \texttt{OrdMonReg} \citep{OrdMonReg}.
Computation relies on a suitable functional $M$ defined on the sets $A \subseteq \{1,\ldots, n\}$ which
generalizes the function $Av$ in \eqref{eq: minmax1}. This functional for the bounded monotone regression in
\eqref{BoundedRegab} is given by \bea M(A)= \Bl(Av(A) \vee \max_A a_L \Br) \wedge \min_A a_U \eea where $\min_A
v = \min_{i \in A} v_i$ and $\max_A v = \max_{i \in A} v_i$. Compare \citet[p. 57]{4b_72}, where a functional
notation is used. However, in the latter reference no formal justification was given for the form of the
functional $M$ nor for the validity of (the modified version of) the PAVA, see the discussion after
Theorem~\ref{min-max}.

\cite{chakra_89} discusses the bounded isotonic regression problem for the
absolute value criterion function, yielding the bounded isotonic median regressor. He proposes a
PAVA-like algorithm as well, and establishes some connections to linear programming theory. Unbounded isotonic
median regression was first considered by \cite{robertson_68}, who provided a min-max formula for the estimator
and a PAVA-like algorithm to compute it. They also studied its consistency.

Now suppose that instead of having only one set of observations $y_1, \ldots, y_n$ at the design
points $x_1, \ldots, x_n$, we are interested in analyzing two sets of data $y_1, \ldots, y_n$
and $z_1, \ldots, z_n$ observed at the same design points. Furthermore, if we have the information that the
underlying true regression curves are increasing and ordered, it is natural to try to construct
estimators that fulfill the same constraints.

The current paper presents a solution to this problem of estimating two isotonic regression curves under the
additional constraint that they are ordered.  This solution is the unique minimizer $(a^*, b^*)$ of the
criterion
\bean
L_2(a, b)&=& \sum_{i=1}^n w_{1,i} (y_i - a_i)^2 + \sum_{i=1}^n w_{2,i} (z_i - b_i)^2 \label{L2}
\eean
over the class of pairs of vectors $(a,b) \in \mathbb{R}^n \times \mathbb{R}^n$ such that $a$ and $b$ are
increasing and $a \le b$, with $w_1$ and $w_2$
given vectors of positive weights in $\mathbb{R}^n$.

\begin{figure}[!h]
\centerline{\epsfig{file = 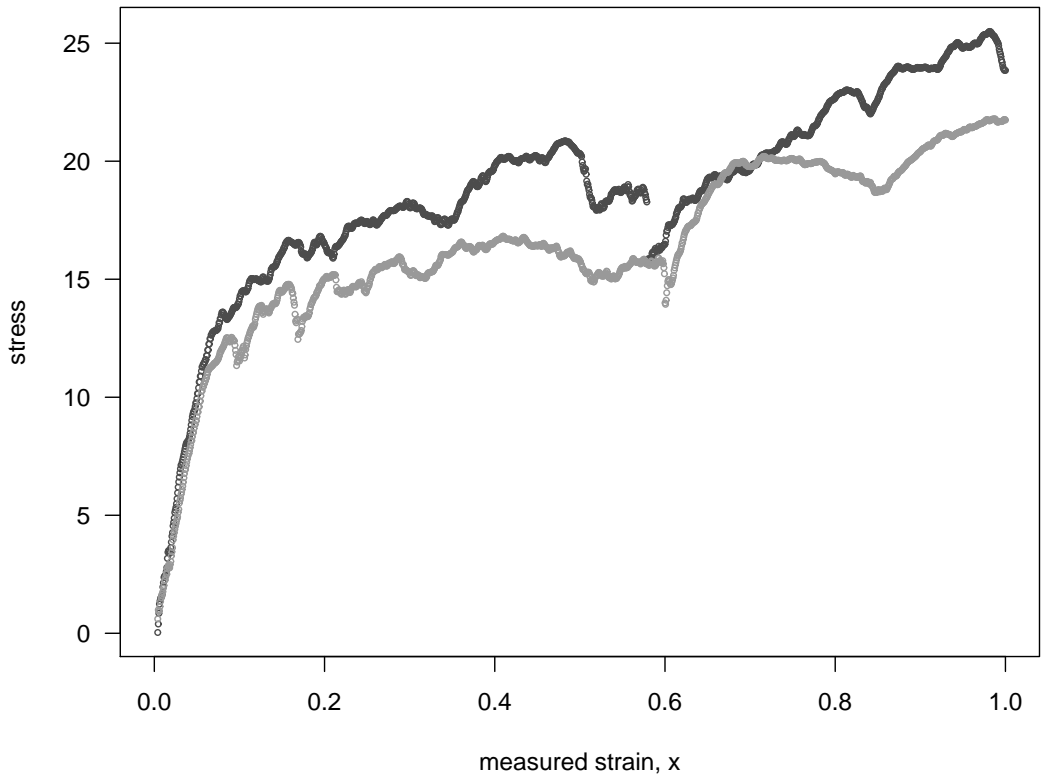, width = 14cm}} \vspace*{-0.5cm} \caption{Original observations.}
\label{fig: meching0}
\end{figure}

The problem was motivated by an application from mechanical engineering. We will make use of experimental data
obtained from dynamic material tests (see \citealp{shim_09}) to illustrate our estimation method. In engineering
mechanics, it is common practice to determine the deformation resistance and strength of materials from uniaxial compression tests at different loading velocities. The experimental results are the so-called stress-strain curves (see Figure~\ref{fig: meching0}),  and these may be used to determine the deformation resistance as a function of the applied deformation. The recorded signals contain substantial noise which is mostly due to variations in the loading velocity and electrical noise in the data acquisition system.

The data in this example consist of 1495 distinct pairs $(x_i, y_i)$ and $(x_i, z_i)$ where $x_i$ is the measured strain, while $y_i$ (gray curve) and $z_i$ (black curve) correspond to the experimental stress
results for two different loading velocities. The true regression curves are expected to be (a) monotone increasing as the stress is known to be an increasing function of the strain (for a given constant loading velocity), and (b) ordered as the deformation resistance typically increases as the loading velocity increases. In Section \ref{sec: algos}, we show the resulting estimates as well as a smoothed version thereof.

We will show that minimizing $L_2$ is equivalent to minimizing another convex functional over the class of
isotonic vectors $a \in \mathbb{R}^n$.  By doing so, we reduce a two-curve problem under the constraints of
monotonicity and ordering to a one-curve problem under the constraint of monotonicity and boundedness. Actually,
we can even perform the minimization over the class of isotonic vectors $(a_1, \ldots, a_{n-1})$  of dimension
$n-1$ satisfying the constraint $a_1 \le \ldots\le a_{n-1} \le a^*_n$ as we can explicitly determine $a^*_n$ by
a \textit{generalized} min-max formula (see Proposition \ref{Expa1}). The solution of this equivalent
minimization problem, which gives the solution $a^*$ (and also $b^*$ because it is a function of $a^*$), is
computed using a projected subgradient algorithm where the projection step is performed using a suitable
generalization of the PAVA. Alternatively, the solution can be computed using Dykstra's algorithm
\citep{dykstra_83}. This point will be further discussed in Section \ref{sec: algos}.

We would like to note that \cite{brunk_66} considered a related problem, that of nonparametric Maximum
likelihood estimation of two ordered cumulative distribution functions. In the same class of problems,
\cite{dykstra_82} treated estimation of survival functions of two stochastically ordered random
variables in the presence of censoring, which was extended by \cite{felz_85} to  $N \ge 2$ stochastically
ordered random variables. The theoretical solution can be related to the well-known Kaplan-Meier estimator and
can be computed using an iterative algorithmic procedure for $N \ge 3$ (see \citealp[p. 1016]{felz_85}). The $\sqrt{n}-$ asymptotics of the estimators for $N=2$, whether there is censoring or not, were established by \cite{praestg_96}.

The paper is organized as follows. In Section~\ref{ordantiestim}, we give the characterization of the ordered
isotonic estimates. We also provide the explicit form of the solution of the related bounded isotonic
regression problem where the upper of the two isotonic curves is assumed to be fully known.

In Section~\ref{sec: algos} we describe the projected subgradient algorithm that we use to compute the Least
Squares estimators of the ordered isotonic regression curves, discuss the connection to  Dykstra's
algorithm \citep{dykstra_83}, and apply the method to real data from mechanical engineering. The technical
proofs are deferred to appendices \ref{proofs} and \ref{subgradient}.

\section{Estimation of two ordered  isotonic regression curves}\label{ordantiestim}

If the larger of the two isotonic curves was known, then there would of course be no need to estimate it.
If we put $a_U = a^0$,
the weighted Least Squares estimate $a^*$ of the smaller isotonic curve is the minimizer of \bea
L(a) &=&  \sum_{i=1}^n w_i (y_i - a_i)^2,
\eea
where $w \in \mathbb{R}^n$ is a vector of given positive weights, and $a \in \II^{a^0}_n$, the class of isotonic vectors $a \in \mathbb{R}^n$ such that $a \le a^0$ and $a^0 \in \R^n$. When the components of $a^0$ are all equal, the vector $a^0$ will be assimilated with the common value of its components as done in Proposition \ref{ASA1} below.

The notation $\II^{w}_n$ will be used again hereafter to denote the class of isotonic vectors $v \in \mathbb{R}^n$ such that $v \le w$.

The statement of \citet[p. 57]{4b_72} implies that if we define
\bea
M(A) = Av(A) \wedge \min_A a^0
\eea
for a subset $A \subseteq \{1, \ldots, n\}$, then the solution $a^*$ can be computed using an
appropriately modified version of the PAVA.
\medskip

\begin{theorem}\label{min-max}
For $i =1,\ldots,n$, we have
\bea
a^*_i = \max_{s \le i} \min_{t \ge i} M(\{s,\ldots,t\}) = \max_{s \le i} \min_{t \ge i} \Bl( Av(\{s,\ldots, t\}) \wedge a^0_s \Br).
\eea
\end{theorem}
To keep this paper at a reasonable length, the proof of Theorem~\ref{min-max} is omitted.
A short note containing a more thorough discussion of the one-curve problem and a proof
of Theorem~\ref{min-max} can be obtained from the authors upon request.
A general description of the modified PAVA and a proof that it works whenever the functional $M$ satisfies the so-called \textit{Averaging Property} can be found in Section \ref{sec: algos}.

We now return to the main subject of this paper.  Theorem \ref{min-max} is crucial for finding the Least Squares estimates of two ordered isotonic regression curves. In particular, the result will be used to develop an appropriate algorithm to compute the solution.

Let $y_1, \ldots, y_n $ and  $z_1, \ldots, z_n $ be the observed data from two unknown isotonic curves $g^\circ_1$ and
$g^\circ_2$ such that $g^\circ_1 \le g^\circ_2$. Given two vectors in $\mathbb{R}^n$ of positive weights $w_1$ and
$w_2$, we would like to minimize \eqref{L2}
over the class of pairs of vectors $(a, b) \in \mathbb{R}^n \times \mathbb{R}^n$ such that $a$ and $b$ are isotonic and $a \le b$. Call this class $\II_n$.

\paragraph*{Existence and uniqueness of the solution.} They follow from convexity and closedness of $\II_n$ and strict convexity of $L_2$.

\paragraph*{Characterization of the solution.} For completeness, we give the characterization of the solution of
minimizing \eqref{L2} over $\II_n$; i.e, a necessary and sufficient condition for  $(a, b) \in \II_n$ to be equal
to this solution. Let $i_1 <  \ldots< i_k $ such that $i_1 = 1, i_k =n$ and
$$
a^*_1 =\ldots=a^*_{i_1} < a^*_{i_1+1} =\ldots = a^*_{i_2-1} < \ldots < a^*_{i_k} =\ldots= a^*_n.
$$
We call $B^0_{i_j}$ (resp. $B^1_{i_j}$) a set of indices $\{i_j, \ldots, i_{j+1}-1 \}, j=1, \ldots, k-1$ such that $a^*_{i_j}= b^*_{i_j}$ (resp. $a^*_{i_j} < b^*_{i_j}$). Similarly, let $l_1 <  \ldots <l_r$ such that $l_1 = 1, l_r = n$ such that
$$
b^*_1 =\ldots=b^*_{l_1} < b^*_{l_1+1} =\ldots = b^*_{l_2-1} < \ldots < b^*_{l_k} =\ldots=b^*_n
$$
and call $C^0_{l_j}$ (resp. $C^1_{l_j}$) a set of indices $\{l_j, \ldots, l_{j+1}-1 \}, j=1, \ldots, r-1$ such that $b^*_{l_j}= a^*_{l_j}$ (resp. $b^*_{l_j} > a^*_{l_j}$).
\begin{theorem}\label{charac2}
The pair $(a^*, b^*) \in \II_n$ is the minimizer of \eqref{L2} if and only if
\bean
\sum_{i=1}^n (a^*_i - y_i)(a^*_i - a_i) w_{1,i}  + \sum_{i=1}^n (b^*_i - z_i)(b^*_i - b_i) w_{2,i} &\ge & 0,  \ \ \forall \ (a, b) \in \II_n  \label{Ineq2} \\
\sum_{s \in \cup_j B^1_{i_j}} (a^*_s - y_s) a^*_s w_{1,s} &= & 0 \label{Eq2}, \ \textrm{and} \\
\sum_{s \in \cup_j C^1_{l_j}} (b^*_s - z_s) b^*_s w_{2,s} &= & 0. \label{Eq3}
\eean

\end{theorem}

\medskip

\pr{} See Appendix~\ref{proofs}.

An explicit formula in the sense of a min-max representation similar to \eqref{eq: minmax1} of $(a^*, b^*)$ turned out
be to hard to find. However, since $a^*$ (resp. $b^*$) is also the
minimizer of
\bea
    \sum_{i=1}^n  (a - y_i)^2 w_{1,i} \ \ \ \Bl(\text{resp.} \sum_{i=1}^n  (b - z_i)^2 w_{2,i}\Br)
\eea over the class $\II^{b^*}_n$
(resp. the class of isotonic vectors $b \in \mathbb{R}^n$ such that $b \ge a^*$), Theorem
\ref{min-max} implies that
\begin{eqnarray}
a^*_i & = &  \max_{s \le i} \min_{t \ge i}  \ (Av_1(\{s,\ldots,t\}) \wedge b^*_s) \label{MinMax1}\\
b^*_i & = & \max_{s \le i} \min_{t \ge i} \ (Av_2(\{s,\ldots,t\}) \vee a^*_t)\label{MinMax2}
\end{eqnarray}
for  $i =1, \ldots, n$, where
\bea
Av_1(A)  =  \frac{\sum_{i \in A} y_i w_{1,i} }{\sum_{i \in A} w_{1,i}}, \ \textrm{and} \  Av_2(A) & = &  \frac{\sum_{i \in A} z_i w_{2,i} }{\sum_{i \in A} w_{2,i}}
\eea
for $A \subseteq \{1, \ldots,  n\}$.

Thus, the solution $(a^*, b^*)$ is a fixed point of the operator $\mathit P:
\II_n \to \II_n$ defined as
\begin{eqnarray}
\mathit P((a,b)) & =&  (\mathit P_1(b), \mathit P_2(a))\label{Operator} \\
                     &= & \left(\max_{s \le i} \min_{t \ge i}  \
(Av_1(\{s,\ldots,t\}) \wedge b_s), \max_{s \le i} \min_{t \ge i} \ (Av_2(\{s,\ldots,t\}) \vee a_t) \right). \nonumber
\end{eqnarray}

However, this fixed point problem does not admit a unique solution. Therefore, there is no guarantee that an
algorithm based on the above min-max formulas yields the solution, except in the unrealistic and uninteresting
case where the starting point of the algorithm is the solution itself. To see that $\mathit P$ does not admit a
unique fixed point, note that the minimizer of the criterion
\bea
\sum_{i=1}^n (a_i - y_i)^2 w_{1,i} +
B \sum_{i=1}^n (b_i - z_i)^2 w_{2,i}
\eea
is a fixed point of $\mathit P$ for any $B > 0$. Therefore, a
computational method based on starting from an initial candidate and then alternating between \eqref{MinMax1}
and \eqref{MinMax2} cannot be successful. In parallel, we have invested a substantial effort in
trying to get a closed form for the estimators. Although we did not succeed, we were able to obtain a closed
form for $a^*_1$ (and by symmetry for $b^*_n$).

\medskip

\begin{proposition}\label{Expa1}
We have that
\bea\label{Expressa1}
a^*_1 =  \min_{t \ge 1} Av_1(\{1,\ldots, t\}) \wedge \min_{t \ge t' \ge 1} \tilde{M}(\{1,\ldots, t\}, \{1,\ldots, t'\})
\eea
where
\bea
 \tilde{M}(A, B) = \frac{Av_1(A) (\sum_{i \in A} w_{1,i}) + Av_2(B) (\sum_{j \in B} w_{2,j})}{\sum_{i \in A} w_{1,i} + \sum_{j \in B} w_{2,j}}.
\eea
By symmetry, we also have that
\begin{eqnarray}\label{Expressbn}
b^*_n =  \max_{t \le n } Av_2(\{t, \ldots,n\}) \vee \max_{t \le t' \le n} \tilde{M}(\{t',\ldots, n\}, \{t,\ldots, n\}).
 \end{eqnarray}
\end{proposition}

\medskip

Some remarks are in order. The expressions obtained above indicate that the Least Squares estimator must depend, as
expected, on the relative ratio of the weights $w_1$ and $w_2$. In particular, if $w_2 =0$ (resp. $w_1 =0$), the
expression of $a^*_1$ (resp. $b^*_n$) specializes to the well-known min-max formula in the classical Least Squares
estimation of an (unbounded) isotonic curve. The expression of $b^*_n$ is essential for our subgradient algorithm below.

\medskip

\par \noindent \textit{Proof of Proposition \ref{Expa1}.} \  See Appendix~\ref{proofs}.

\medskip

In the next section, we describe how we can make use of the min-max formula in \eqref{MinMax1} to compute the estimators using a projected subgradient algorithm. As mentioned above,  we use in this algorithm the identity \eqref{Expressbn} given in the previous proposition.

\section{Algorithms and Application to real data} \label{sec: algos}

In this section, we show that the bounded isotonic estimator can be computed using the well-known PAVA, or to be
more precise a modified version of it. Recall that the bounded isotonic estimator in the one-curve problem is
given by \bea a^*_i &=& \max_{s\le i} \min_{t \ge i} M(\{s,\ldots, t\}) \eea where $M(A) = Av(A) \vee \max_A
a^0$ for any $ A \subseteq \{1, \ldots, n\}$. That $a^*$ can be computed using a PAVA is a consequence of a more general
result. Namely, that a functional $M$ of sets $A \subseteq \{1, \ldots, n\}$
satisfies what is referred to as the \textit{\ap,} (see \citealp[p. 138]{chakra_89}), also called \textit{Cauchy
Mean Value Property} by \citet[Section 1]{leurgans_81}. See also \citet[p. 390]{robertson_88}. Note that in the
classical unconstrained monotone regression problem, the min-max expression of the Least Squares estimator
follows from Theorem 2.8 in \citet[p. 80]{4b_72}.

\subsection{Getting the min-max solution by the PAVA}


First, let us describe how the PAVA works for some set functional $M$.

\begin{itemize}
\item At every step the current configuration is given by a subdivision of $\{1,\ldots,n\}$ into $k$ subsets $S_1=\{1,\ldots,i_1\},\,S_2=\{i_1+1,\ldots,i_2\},\dots ,\,S_k=\{i_{k-1}+1,\ldots, n\}$ for some indices $1=i_0 \leq i_1<i_2<\dots <i_{k-1}<i_k=n$.
\item The initial configuration is given by the finest subdivision; i.e., $I_j=\{j\}$.
\item At every step we look at the values of $M$ on the sets of the subdivision. A violation is noted each time there exists a value $j$ such that $M(S_j) > M(S_{j+1})$. We consider the first violation (the one corresponding to the smallest $j$) and then merge the subsets $S_j$ and $S_{j+1}$ into one interval.
\item Given a new subdivision (which has one subset less than the previous one), we look for possible violations.
\item The algorithm stops when there are no violations left.
\end{itemize}
Since for any violation a merging is performed (thus reducing the number of subsets), it is clear that the algorithm stops after a finite number of iterations. 

We require now the set functional $M$ to satisfy the following property. See \citet[Section 1]{leurgans_81},
\citet[p. 390]{robertson_88} and \citet[p. 138]{chakra_89}.

\begin{definition}\label{def: M}
We say that the functional $M$ satisfies the \ap if for any sets $A$ and $B$ such that $A\cap B=\emptyset$ we
have that
\bea
    \min \{ M(A), M(B)\}\leq M(A\cup B)\leq \max \{ M(A), M(B)\}.
\eea
\end{definition}

\medskip

\noindent If $h$ and $w > 0$ are given vectors $\in \mathbb R^n$, then beside
\bea
A \mapsto Av(A) &=& \sum_{i \in A} w_i h_i / \sum_{i \in  A} w_i,
\eea
the following examples of functions also satisfy the \ap:
\bea
    A &\mapsto & \Bl(Av(A) \vee \max_A h^1_i\Br) \wedge \min_A h^0, \ \ \textrm{with $h^0, h^1$ two vectors $\in \mathbb R^n$}, \\
    A &\mapsto & \min_{A} h = \min_{i \in A} h_i, \\
    A&\mapsto &  \mathrm{med}_{A} \ h = \argmin_{m \in \mathbb R} \sum_{i \in A} \vert h_i - m \vert w_i\\
    && \ \textrm{where the $\argmin$ is taken to be the smallest $m$ in case non-uniqueness occurs}, \\
    A &\mapsto & \max_{A} h = \max_{i \in A} h_i.
\eea

Note that the maximum, the minimum and the sum of two functionals satisfying the \ap satisfy the same property as well.

\medskip

\begin{theorem}\label{PAVA}
The final configuration obtained by the PAVA is such that the two following properties are satisfied.
\begin{enumerate}
\item The functional $M$ is increasing on the sets of the subdivision.
\item If one of the sets $S_j=C\cup D$ is the disjoint union of two subsets $C=\{i_{j-1}+1,\ldots, k\}$ and $D=\{k+1,\ldots, i_{j}\}$, then $M(C)> M(D)$; i.e., a finer subdivision would necessarily cause a violation.
\end{enumerate}
\end{theorem}

\pr{} The fact that $M$ is increasing on the final configuration is an easy consequence of the absence of violations (otherwise the algorithm would not have stopped).

As for the second part of the property, note that this is satisfied by the initial configuration (since no set is the disjoint union of two non-trivial subsets), as well as by any configuration that one could obtain after the first merging (since a merging occurs only because of a violation). Now we will use an inductive reasoning.

To this end, we have to check two situations: Suppose we merge two subsequent sets $A$ and $B$ and want to check whether there is a violation on $C$ and $D$, with $A\cup B=C\cup D$. We are in one of the two following cases: either $A=A_1\cup A_2$, $C=A_1$ and $D=A_2\cup B$, or $B=B_1\cup B_2$, $C=A\cup B_1$ and $D=B_2$ (the case $C=A$ and $D=B$ is trivial).

In the first case, if we suppose $M(D)\geq M(C)$, we get
$$M(A_2\cup B)\geq M(A_1),\;M(A_2)<M(A_1),\;M(B)<M(A)=M(A_1\cup A_2),$$
(the first inequality follows by assumption, the second by induction, and the third is true since $A$ and $B$ have been merged) and this is impossible since one would conclude that
$$\max\{M(A_2),M(B)\}\geq M(A_1) > M(A_2),$$
and hence $M(A)> M(B)\geq M(A_1)>M(A_2)$, which implies $M(A)>\max\{M(A_1),M(A_2)\}$, which contradicts the \ap.

In the second case we would have
$$M(A\cup B_1)\leq M(B_2),\;M(B_2)<M(B_1),\;M(A)> M(B)=M(B_1\cup B_2),$$
which implies
$$\min\{M(A),M(B_1)\}\leq M(B_2)<M(B_1),$$
and then $\min\{M(A),M(B_1)\}=M(A)$ and $M(A)\le M(B_2)<M(B_1)$, which contradicts either $M(A)<M(B)$ or the \ap.   \hfill $\Box$

\medskip

\begin{theorem}\label{PAVA2}
If $(S_j)_j$ is the partition  obtained at the end of the PAVA described above, then $m_i=M(S_{j_i})$ such that $ i\in S_{j_i}$ takes the same values given by the min-max formula for the index $i$.
\end{theorem}

\pr{} See Appendix~\ref{proofs}.

\subsection{Shor's projected subgradient and Dykstra's iterative cyclic projection algorithm}

The minimization problem considered in this paper can be easily recognized as a projection problem onto the intersection of the three following closed convex cones in $\mathbb R^n \times \mathbb R^n$
\bea
\{(a,b) : a \mbox{ is increasing} \}, \  \{(a,b) : b \mbox{ is increasing} \}, \ \textrm{and} \ \{(a,b): a\leq b\}.
\eea
Projections onto the first two cones can be computed by PAVA, and onto the last one by replacing the components of each pair $(a_i,b_i)$ violating the constraint (i.e. $a_i>b_i$) by the
weighted average $(w_{1,i} a_i + w_{2,i} b_i)/(w_{1,i} + w_{2,i}) $ of $a_i$ and $b_i$. Implementation of Dykstra's
algorithm \citep{dykstra_83} is then straightforward.

Yet, our algorithm has preferable features as we will now explain. The algorithm developped by Dykstra is
well-suited for projections onto intersections of convex sets 
or half-spaces (see \citealp{bregman_03}), while the algorithm we propose can handle a larger class of
minimization problems which involve the set of isotonic vectors, and are not necessarily projections.  For instance, simple modifications of our algorithm  would allow us to minimize any objective function of the form
\bea
(a, b) \mapsto F(a, w_1)+\sum_{i=1}^n w_{2,i} (z_i-b_i)^2
\eea
under the same constraints on $a$ and $b$, where $F$ is any convex and differentiable function. The second
quadratic term can be also replaced by a different penalization term depending e.g. on an $L_p$-distance. Indeed, it suffices to modify the computations involved in the PAVA by adapting them to various functionals satisfying the Averaging Property (see Section \ref{sec: algos}).

Our algorithm is easy to understand and is only based on a classical gradient
method. Once the minimization is performed with respect to one of the variables, the objective
function with respect to the remaining variable is still explicit, but no more differentiable.  This is the main reason for which the algorithm
is actually a subgradient descent. We believe that the explicit nature of the computations in our subgradient algorithm are exactly
the key feature for the possibility of understanding and/or modifying it.

However, we would like to point out the merits of Dykstra's algorithm in this specific setting. Since  it is tailored
for a Least Squares problem, and because only three very simple projection cones are involved,
Dykstra's algorithm (see below for details) computes the minimum of the criterion $L_2$ given in \eqref{L2} faster than the subgradient algorithm,
although Dykstra's algorithm is typically considered to be rather slow (see e.g. \citealp{mammen_91b} or \citealp{birke_07}).
Note that the choice of the stopping criterion in this algorithm may be delicate, see \cite{birgin_05}. However, this was not an issue in our setting.

\subsection{Preparing for a projected subgradient algorithm}

The following proposition is crucial for computing the ordered isotonic estimators via a projected subgradient algorithm.

\begin{proposition}\label{ASA1} Let $\Psi$ be the criterion
\begin{eqnarray}
\Psi(b_1, \ldots, b_{n-1}) &=& \sum_{i=1}^{n} \Big(\max_{s \le i} (G_{s,i} \wedge b_s) - y_i\Big)^2 w_{1,i} + \sum_{i=1}^{n-1} (b_i - z_i)^2 w_{2,i}  \label{Psi}
\end{eqnarray}
which is to be minimized on the convex set
\bea
\II^{b^*_n}_{n-1} = \{(b_1, \ldots, b_{n-1}) \in \R^{n-1}: \ b_1 \le b_2 \le \ldots \le b_{n-1} \le b^*_n  \}
\eea
where
\bea
G_{s,i} = \min_{t \ge i} Av_1(\{s,\ldots, t\}) \ \ \textrm{and \ $b_n = b^*_n$ in } \eqref{Psi}.
\eea

The criterion $\Psi$ is convex. Furthermore, its unique minimizer $(b^{**}_1, \ldots, b^{**}_{n-1})$ equals $(b^*_1, \ldots, b^*_{n-1})$.
\end{proposition}

\medskip

\pr{} Let us write
\bea
\II = \II^\infty_n = \{ a= (a_1, \ldots, a_n) \in \mathbb R^n: a_1 \le \ldots\le a_n\},
\eea
\bea
\II^*_n = \Big \{ b = (b_1, \ldots, b_n): (b_1, \ldots, b_{n-1}) \in \II^{b_n^*}_{n-1} \ \textrm{and} \ b_n = b^*_n \Big \}
\eea
and  consider
\bea
\II^{b}_n = \{ a:  a \in \II  \ \textrm{and} \ a \le b \}
\eea
for $b \in \II^*_n$.

Now note that the min-max formula in \eqref{MinMax1} allows us to write
\bea
&& \sum_{j=1}^{n} \Big(\max_{s \le j} (G_{s,j} \wedge b_s) - y_j\Big)^2 w_{1,j} + \sum_{j=1}^{n-1} (b_j - z_j)^2 w_{2,j}\\
 && =  \min_{ a \in \II^b_n} \sum_{j=1}^n (a_j - y_j)^2 w_{1,j} +  \sum_{j=1}^{n-1} (b_j - z_j)^2 w_{2,j}.
\eea
Hence, we have for $b \in \II^*_n $
\bea
\Psi(b_1, \ldots, b_{n-1}) &= & \min_{ a \in  \II^b_n} \sum_{j=1}^n (a_j - y_j)^2 w_{1,j} +  \sum_{j=1}^{n-1} (b_j - z_j)^2 w_{2,j} \\
& = & \sum_{j=1}^n (\tilde{a}_j(b) - y_j)^2 w_{1,j} +  \sum_{j=1}^{n-1} (b_j - z_j)^2 w_{2,j}
\eea
where $\tilde{a}_j(b) = \max_{s \le j} (G_{s,j} \wedge b_s)$ is the $j$-th component of the minimizer of the function $\sum_{j=1}^n (a_j - y_j)^2 w_{1,j}$ in $\II^b_n$. Let $\lambda \in [0,1]$, and $b$ and $b'$ in $\II^*_n$.  By definition of $\II^b_n$ and $\II^{b'}_n$, we have that
\bea
\lambda \ \tilde{a}(b) + (1-\lambda) \ \tilde{a}(b') \le \lambda \ b + (1-\lambda) \ b'
\eea
and hence
\bea
&&\sum_{j=1}^n \Big(\tilde{a}_j(\lambda \ b + (1-\lambda) \ b') - y_j \Big)^2 w_{1,j} \\
&& \le
\sum_{j=1}^n \Big(\lambda \ \tilde{a}(b) + (1-\lambda) \ \tilde{a}(b') - y_j\Big)^2 w_{1,j} \\
&& \le  \lambda \sum_{j=1}^n \ \Big(\tilde{a}_j(b)- y_j \Big)^2 w_{1,j} +(1-\lambda) \sum_{j=1}^n \ \Big(\tilde{a}_j(b')- y_j \Big)^2 w_{1,j}.
\eea

This shows convexity of the first term of $\Psi$.  Convexity of $\Psi$ now follows from convexity of the function
$\sum_{j=1}^{n-1} (b_j - z_j)^2 w_{2,j}$ and the fact that the sum of two convex functions defined on the same domain is also convex. \hfill  $\Box$

The idea behind considering the convex functional $\Psi$ is to reduce the dimensionality of the problem as well as the number of constraints (from $3n-2$ to $n-1$ constraints).  Once $\Psi$ is minimized; i.e, the isotonic estimate $b^*$ is computed, $a^*$ can be obtained using the min-max formula given in  \eqref{MinMax1}. However, the convex functional $\Psi$ is not continuously differentiable, hence the need for an optimization algorithm that uses the subgradient instead of the gradient as the latter is not defined everywhere.

\subsection{A projected subgradient algorithm to compute $b_1^*, \ldots, b_{n-1}^*$}

To minimize the non-smooth convex function $\Psi$ we use a projected subgradient algorithm. Since the gradient
does not exist on the entire domain of the function, one has to resort to computation of a subgradient, the
analogue of the gradient at points where the latter does not exist. As opposed to classical methods developed
for minimizing smooth functions, the procedure of searching for the direction of descent and steplengths is
entirely different. The classical reference for subgradient algorithms is \cite{shor_85}. \cite{boyd_03}
provide a nice summary of the topic, including the projected variant. Note that a recent application in
statistics of the subgradient algorithms gives now the possibility to compute the log-concave density estimator
in high dimensions; see \cite{cule_08}.

\paragraph*{The main steps of the algorithm.}  Now recall that the functional $\Psi$ should be minimized over the $(n-1)-$ dimensional convex set $\II^{b_n^*}_{n-1}$ given in Proposition~\ref{ASA1}. Of course, this is the same as minimizing $\Psi$ over the $n-$ dimensional convex set $\{(b_1, \ldots, b_n) \ | \ b_1 \le \ldots \le b_{n-1} \}$,
starting with an initial vector $(b^{(0)}_1, \ldots, b^{(0)}_n)$ such that $b^{(0)}_n = b^*_n$ and constraining the $n-$th component of the sub-gradient of $\Psi$ to be equal to 0.

Given a steplength $\tau_k$, the new iterate  $\ve{b}^{k+1} = (b_1^k, \ldots, b_{n}^k)$ at the $k-$th iteration of  a subgradient algorithm is given by
\bea
    \ve{v}_{k+1} &=& \ve{b}_k - \tau_k \ve{D}_k,
\eea
where $\ve{D}_k$ is the subgradient calculated at the previous iterate; i.e., $\ve{D}_k = \tilde{\nabla}\Psi(\ve{v}_k)$ (see Appendix~\ref{subgradient}).   However, it may happen that $\ve{v}_{k+1}$ is not admissible; i.e. $(b^{k+1}_1, \ldots, b^{k+1}_{n-1})$ does not belong to $\II^{b_n^*}_{n-1}$. When this occurs, an $L_2$ projection of this iterate onto $\II^{b_n^*}_{n-1}$ is performed. This is equivalent to finding the minimizer of
\bea
\sum_{i=1}^n  (a_i - b^{k+1}_i)^2
\eea
over the set $\II^{b^*_n}_n$. The latter problem can be solved using the generalized PAVA
for bounded isotonic regression as described above.

The computation of the subgradient $\ve{D}_k$ is described in detail in Appendix~\ref{subgradient}. As for the steplength $\tau_k$,
we start the algorithm with a constant steplength. Once a pre-specified number of iterations has been reached we switch to
\bea
    \tau_{k+1} &=& (h_k^{0.1} \Vert \ve{D}_k \Vert_2)^{-1}
\eea where $\gamma_k := h_k^{-0.1}$ is such that $0 \le \gamma_k \to 0$ as $k \to \infty$ and
$\sum_{k=1}^\infty \gamma_{k} = \infty$. Here, $\Vert \cdot \Vert_2$ denotes the $L_2$-norm of
a vector in $\R^n$. This combination of constant and non-summable diminishing steplength showed a good performance in our implementation of the algorithm over other classical choices of $(\gamma_k)_k$. Furthermore, convergence is ensured by the following theorem.

\begin{theorem}\textbf{(\cite{boyd_03})}\label{algo conv}
A subgradient algorithm complemented with least-square projection and using non-summable diminishing steplength
yields for any $\eta > 0$ after $k = k(\eta)$ iterations a vector $b^k := (b_1^k, \ldots, b_{n}^k)$ such
that \bea
    \min_{i=1,\ldots, k}\Psi(b^i) - \Psi(b^*) &\le& \eta,
\eea where $b^* = (b_1^*, \ldots, b_{n}^*)$ is the vector given in Proposition~\ref{ASA1}.
\end{theorem}

The proof can be found in \cite{boyd_03} by combining their arguments in Sections 2 and 3. Note that in our implementation we do not keep track of the iterate that yielded the minimal value of $\Psi$, since we apply a problem-motivated stopping criterion that guarantees us to have reached an iterate that is sufficiently close to $b^* = (b_1^*, \ldots, b_n^*)$.

\paragraph*{Choice of stopping rule.}
Since in subgradient algorithms the convex target functional does not necessarily monotonically decrease with increasing
number of iterations, the choice of a suitable stopping criterion is delicate. However, in our specific setting
we use the fact that $(a^*, b^*)$ is a fixed point of the operator $\mathit  P $ defined in \eqref{Operator} where  $ a^* = \mathit P_1(b^*)$; the solution of \eqref{L} with upper bound $b^*$. This motivates iterating the algorithm until the  difference of entries of the two vectors
$b^k$ and $b_\#^k$ where
$$
b_\#^k =  \mathit P_2 \circ \mathit P_1(b^k)
$$
is below a pre-specified positive constant $\delta$.

\paragraph*{The implementation.} The Dykstra and the projected subgradient algorithms  as well as
the generalized PAVA for computing the solution in the one curve problem under the constraints in \eqref{BoundedRegab} were all implemented in \texttt{R}
\citep{R}. The corresponding package \texttt{OrdMonReg}
\cite{OrdMonReg} is available on CRAN. Note that the data analyzed in Section~\ref{sec: real data} is made
available as a dataset in \texttt{OrdMonReg}.

To conclude this section on the algorithmic aspects of our work, we would like to mention the work by \cite{beran_09}
who propose an active set algorithm which can be tailored to solve the problem given in \eqref{L2} for an
arbitrary number of ordered monotone curves. However, \cite{beran_09} do not provide an analysis of the structure of the estimated curves such as characterizations and rather put their emphasis on the algorithmic developments of the problem.

\subsection{Real data example from mechanical engineering} \label{sec: real data}

We would like to estimate the stress-strain curves based on the available experimental data for two different velocity levels (see Figure~\ref{fig: meching0}). The expected curves have to be isotonic and ordered. The data consist of 1495 pairs $(x_i, y_i)$ and $(x_i, z_i)$.  The values of the measured strain of the material (on the $x$-axis), are actually defined as $(-)$ the logarithm of the ratio of the current over the initial specimen length. The values are positive and take the maximal value 1, which corresponds to a maximum shortening of 63\%.

Furthermore, since the stress measurements for different velocities are not performed exactly at the same strain, the values of the stress have been interpolated at equally spaced values of the strain. As pointed out by a referee, this will induce correlation between the strain data. Even if the strain measurement were not interpolated, having correlated stress measurements is rather inevitable in this particular application because of the data processing procedures associated with the measurement technique (see \citealp{shim_09}). The estimation method is however still applicable.  When studying
statistical properties of the isotonic estimators such as consistency and convergence, the correlation between the data
should, of course, be taken into account.

In such problems, practitioners usually fit parametric models using a trial and error approach in an attempt to capture
monotonicity of the stress-strain curves as well as their ordering. The methods used are rather arbitrary and can
also be time consuming, hence the need for an alternative estimation approach. Our main goal is to provide those
practitioners with a rigorous way for estimating the ordered stress-strain curves.

\begin{figure}[!h]
\centerline{\epsfig{file = 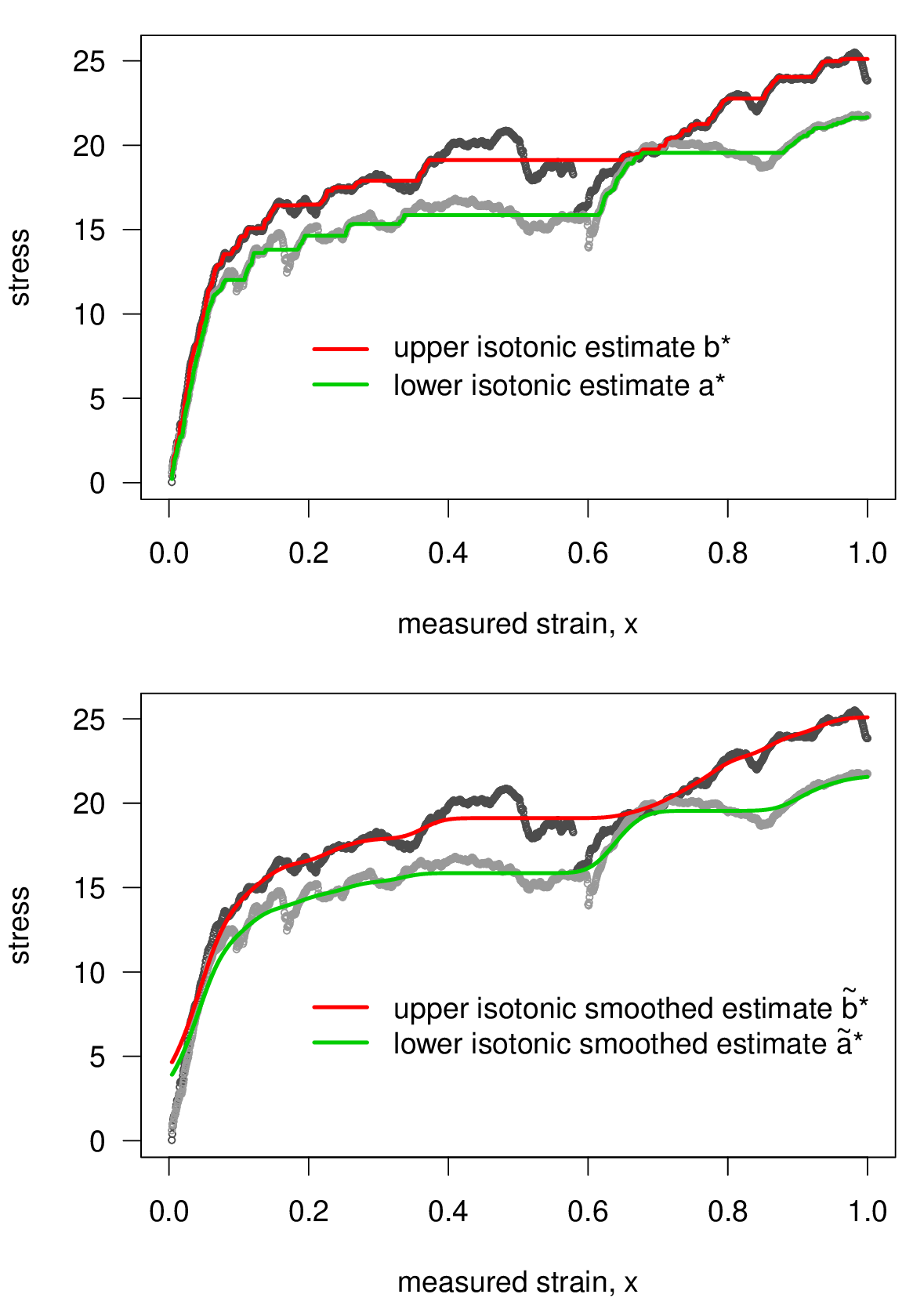, width = 14cm}}
\vspace*{-0.5cm}
\caption{Original observations, isotonic and isotonic smoothed estimates.}
\label{fig: meching1}
\end{figure}

In Figure~\ref{fig: meching1} (upper plot) we provide the original data (black and gray dots) and the proposed ordered isotonic estimates
$a^*$ and $b^*$ as described above. Being step functions, the estimated isotonic curves are
non-smooth, a well known drawback of isotonic regression, see among others \cite{wright_78} and
\cite{mukerjee_86}. The latter author pioneered the combination of isotonization followed by kernel smoothing.
A thorough asymptotic analysis of the smoothed isotonized and the isotonic smooth estimators was given by \cite{mammen_91}. \citet[p. 743]{mukerjee_86} shows that monotonicity of the regression
function is preserved by the smoothing operation if the used kernel is log-concave. Thus, we define our smoothed
ordered monotone estimators by
\bea
    \tilde a_h^*(x) \ = \ \frac{\sum_{i = 1}^n K_h(x - t)a_i^*}{\sum_{i = 1}^n K_h(x - x_i)} && \ \textrm{and} \ \
    \tilde b_h^*(x) \ = \ \frac{\sum_{i = 1}^n K_h(x - t)b_i^*}{\sum_{i = 1}^n K_h(x - x_i)}
\eea
for $0 \le x \le 1 $. For simplicity, we used the kernel $K_h(x) = \phi(x/h)$ where $\phi$
is the density function of a standard normal distribution which is clearly log-concave.
Figure~\ref{fig: meching1} (lower plot) depicts the smoothed isotonic estimates. We set
the bandwidth to $h = 0.1 n^{-1/5} \approx 0.023$.

Motivated by estimation of stress-strain curves, an application from mechanical engineering, we consider in this paper weighted Least Squares estimators in the problem of estimating two ordered isotonic regression curves. We provide characterizations of the solution and describe a projected subgradient
algorithm which can be used to compute this solution. As a by-product, we show how an adaptation of the well-known PAVA can be used to compute min-max estimators for any set functional satisfying the Averaging Property.

\medskip

\section*{Acknowledgements.} The first author would like to thank C\'ecile Durot for some interesting discussions
around the subject. We also thank JongMin Shim for having made the data available to us, a reviewer for drawing
our attention to Dykstra's algorithm, and another reviewer for helpful remarks.

\bigskip


\bibliographystyle{ims}
\bibliography{stat}

\begin{thebibliography}{27}
\expandafter\ifx\csname natexlab\endcsname\relax\def\natexlab#1{#1}\fi
\expandafter\ifx\csname url\endcsname\relax
  \def\url#1{\texttt{#1}}\fi
\expandafter\ifx\csname urlprefix\endcsname\relax\def\urlprefix{URL }\fi

\bibitem[{Balabdaoui et~al.(2009)Balabdaoui, Rufibach and
  Santambrogio}]{OrdMonReg}
\textsc{Balabdaoui, F.}, \textsc{Rufibach, K.} and \textsc{Santambrogio, F.}
  (2009).
\newblock \textit{OrdMonReg: Compute least squares estimates of one bounded or
  two ordered isotonic regression curves}.
\newblock R package version 1.0.2.

\bibitem[{Barlow et~al.(1972)Barlow, Bartholomew, Bremner and Brunk}]{4b_72}
\textsc{Barlow, R.~E.}, \textsc{Bartholomew, D.~J.}, \textsc{Bremner, J.~M.}
  and \textsc{Brunk, H.~D.} (1972).
\newblock \textit{Statistical inference under order restrictions. {T}he theory
  and application of isotonic regression}.
\newblock John Wiley \& Sons, London-New York-Sydney.
\newblock Wiley Series in Probability and Mathematical Statistics.

\bibitem[{Beran and D\"umbgen(2009)}]{beran_09}
\textsc{Beran, R.} and \textsc{D\"umbgen, L.} (2009).
\newblock Least squares and shrinkage estimation under bimonotonicity
  constraints.
\newblock \textit{Statistics and Computing, to appear} .

\bibitem[{Birgin and Raydan(2005)}]{birgin_05}
\textsc{Birgin, E.~G.} and \textsc{Raydan, M.} (2005).
\newblock Robust stopping criteria for {D}ykstra's algorithm.
\newblock \textit{SIAM J. Sci. Comput.} \textbf{26} 1405--1414 (electronic).

\bibitem[{Birke and Dette(2007)}]{birke_07}
\textsc{Birke, M.} and \textsc{Dette, H.} (2007).
\newblock Estimating a convex function in nonparametric regression.
\newblock \textit{Scand. J. Statist.} \textbf{34} 384--404.

\bibitem[{Boyd et~al.(2003)Boyd, Xiao and Mutapcir}]{boyd_03}
\textsc{Boyd, S.}, \textsc{Xiao, L.} and \textsc{Mutapcir, A.} (2003).
\newblock Subgradient methods.
\newblock Lecture Notes, Stanford University.
\newline\urlprefix\url{http://www.stanford.edu/class/ee392o/subgrad_method.pdf}

\bibitem[{Bregman et~al.(2003)Bregman, Censor, Reich and
  Zepkowitz-Malachi}]{bregman_03}
\textsc{Bregman, L.~M.}, \textsc{Censor, Y.}, \textsc{Reich, S.} and
  \textsc{Zepkowitz-Malachi, Y.} (2003).
\newblock Finding the projection of a point onto the intersection of convex
  sets via projections onto half-spaces.
\newblock \textit{J. Approx. Theory} \textbf{124} 194--218.

\bibitem[{Brunk(1958)}]{brunk_58}
\textsc{Brunk, H.~D.} (1958).
\newblock On the estimation of parameters restricted by inequalities.
\newblock \textit{Ann. Math. Statist.} \textbf{29} 437--454.

\bibitem[{Brunk et~al.(1966)Brunk, Franck, Hanson and Hogg}]{brunk_66}
\textsc{Brunk, H.~D.}, \textsc{Franck, W.~E.}, \textsc{Hanson, D.~L.} and
  \textsc{Hogg, R.~V.} (1966).
\newblock Maximum likelihood estimation of the distributions of two
  stochastically ordered random variables.
\newblock \textit{J. Amer. Statist. Assoc.} \textbf{61} 1067--1080.

\bibitem[{Chakravarti(1989)}]{chakra_89}
\textsc{Chakravarti, N.} (1989).
\newblock Bounded isotonic median regression.
\newblock \textit{Comput. Statist. Data Anal.} \textbf{8} 135--142.

\bibitem[{Cule et~al.(2008)Cule, Samworth and Stewart}]{cule_08}
\textsc{Cule, M.}, \textsc{Samworth, R.} and \textsc{Stewart, M.} (2008).
\newblock Maximum likelihood estimation of a multidimensional log-concave
  density.
\newline\urlprefix\url{http://www.citebase.org/abstract?id=oai:arXiv.org:0804.%
3989}

\bibitem[{Dykstra(1982)}]{dykstra_82}
\textsc{Dykstra, R.~L.} (1982).
\newblock Maximum likelihood estimation of the survival functions of
  stochastically ordered random variables.
\newblock \textit{J. Amer. Statist. Assoc.} \textbf{77} 621--628.

\bibitem[{Dykstra(1983)}]{dykstra_83}
\textsc{Dykstra, R.~L.} (1983).
\newblock An algorithm for restricted least squares regression.
\newblock \textit{J. Amer. Statist. Assoc.} \textbf{78} 837--842.

\bibitem[{Feltz and Dykstra(1985)}]{felz_85}
\textsc{Feltz, C.~J.} and \textsc{Dykstra, R.~L.} (1985).
\newblock Maximum likelihood estimation of the survival functions of {$N$}
  stochastically ordered random variables.
\newblock \textit{J. Amer. Statist. Assoc.} \textbf{80} 1012--1019.

\bibitem[{Leurgans(1981)}]{leurgans_81}
\textsc{Leurgans, S.} (1981).
\newblock The {C}auchy mean value property and linear functions of order
  statistics.
\newblock \textit{Ann. Statist.} \textbf{9} 905--908.

\bibitem[{Mammen(1991{\natexlab{a}})}]{mammen_91b}
\textsc{Mammen, E.} (1991{\natexlab{a}}).
\newblock Estimating a smooth monotone regression function.
\newblock \textit{Ann. Statist.} \textbf{19} 724--740.

\bibitem[{Mammen(1991{\natexlab{b}})}]{mammen_91}
\textsc{Mammen, E.} (1991{\natexlab{b}}).
\newblock Estimating a smooth monotone regression function.
\newblock \textit{Ann. Statist.} \textbf{19} 724--740.

\bibitem[{Mukerjee(1988)}]{mukerjee_86}
\textsc{Mukerjee, H.} (1988).
\newblock Monotone nonparameteric regression.
\newblock \textit{Ann. Statist.} \textbf{16} 741--750.

\bibitem[{Pr{\ae}stgaard and Huang(1996)}]{praestg_96}
\textsc{Pr{\ae}stgaard, J.~T.} and \textsc{Huang, J.} (1996).
\newblock Asymptotic theory for nonparametric estimation of survival curves
  under order restrictions.
\newblock \textit{Ann. Statist.} \textbf{24} 1679--1716.

\bibitem[{{R Development Core Team}(2008)}]{R}
\textsc{{R Development Core Team}} (2008).
\newblock \textit{R: A Language and Environment for Statistical Computing}.
\newblock R Foundation for Statistical Computing, Vienna, Austria.
\newblock {ISBN} 3-900051-07-0.
\newline\urlprefix\url{http://www.R-project.org}

\bibitem[{Robertson and Waltman(1968)}]{robertson_68}
\textsc{Robertson, T.} and \textsc{Waltman, P.} (1968).
\newblock On estimating monotone parameters.
\newblock \textit{Ann. Math. Statist} \textbf{39} 1030--1039.

\bibitem[{Robertson et~al.(1988)Robertson, Wright and Dykstra}]{robertson_88}
\textsc{Robertson, T.}, \textsc{Wright, F.~T.} and \textsc{Dykstra, R.~L.}
  (1988).
\newblock \textit{Order restricted statistical inference}.
\newblock Wiley Series in Probability and Mathematical Statistics: Probability
  and Mathematical Statistics, John Wiley \& Sons Ltd., Chichester.

\bibitem[{Shim and Mohr(2009)}]{shim_09}
\textsc{Shim, J.} and \textsc{Mohr, D.} (2009).
\newblock Using split hopkinson pressure bars to perform large strain
  compression tests on polyurea at low, intermediate and high strain rates.
\newblock \textit{International Journal of Impact Engineering} \textbf{36} 1116
  -- 1127.

\bibitem[{Shor(1985)}]{shor_85}
\textsc{Shor, N.} (1985).
\newblock \textit{Minimization Methods for Non-Differentiable Functions}.
\newblock Springer, Berlin.

\bibitem[{van Eeden(1957{\natexlab{a}})}]{eeden_57a}
\textsc{van Eeden, C.} (1957{\natexlab{a}}).
\newblock Maximum likelihood estimation of partially or completely ordered
  parameters. {I}.
\newblock \textit{Nederl. Akad. Wetensch. Proc. Ser. A. {\bf 60} = Indag.
  Math.} \textbf{19} 128--136.

\bibitem[{van Eeden(1957{\natexlab{b}})}]{eeden_57b}
\textsc{van Eeden, C.} (1957{\natexlab{b}}).
\newblock Maximum likelihood estimation of partially or completely ordered
  parameters. {II}.
\newblock \textit{Nederl. Akad. Wetensch. Proc. Ser. A. {\bf 60} = Indag.
  Math.} \textbf{19} 201--211.

\bibitem[{Wright(1978)}]{wright_78}
\textsc{Wright, F.~T.} (1978).
\newblock Estimating strictly increasing regression functions.
\newblock \textit{Journal of the American Statistical Association} \textbf{73}
  636--639.
\newline\urlprefix\url{http://www.jstor.org/stable/2286615}

\end{thebibliography}

\begin{appendix}

\section{Proofs} \label{proofs}

\pr{\ of \ Theorem \ref{charac2}} Suppose that $(a^*, b^*)$ is the solution. For $\epsilon \in (0,1)$, and $(a, b) \in \II_n$  consider the pair $(a^\epsilon, b^\epsilon) \in \mathbb{R}^n \times \mathbb{R}^n$ defined as
\bea
a^\epsilon &= & a^* + \epsilon (a - a^*)    \\
b^\epsilon  & = & b^* + \epsilon (b -b^*).
\eea
For $i \le j \in \{1, \ldots, n\}$, we have
\bea
a^\epsilon_j - a^\epsilon_i & = &  (1-\epsilon) (a^*_j - a^*_i)+  \epsilon (a_j - a_i) \ge 0 \\
b^\epsilon_j - b^\epsilon_i & = &  (1-\epsilon) (b^*_j - b^*_i)+  \epsilon (b_j - b_i) \ge 0.
\eea
Also, for $i \in \{1, \ldots, n\}$ we have
\bea
a^\epsilon_i - b^\epsilon_i & = &  (1-\epsilon) (a^*_i - b^*_i)+  \epsilon (a_i - b_i) \le 0.
\eea
Hence, $(a^\epsilon, b^\epsilon) \in \II_n$, and
\bea
0 &\le & \lim_{\epsilon \searrow 0} \frac{1}{\epsilon} (L_2(a^\epsilon,b^\epsilon) - L_2(a^*,b^*)) \\
  & = &  \sum_{i=1}^n (a^*_i - y_i)(a_i - a^*_i) w_{1,i} + \sum_{i=1}^n (b^*_i - z_i)(b_i - b^*_i) w_{2,i}
\eea
yielding the inequality in \eqref{Ineq2}.

Now consider the vectors $a^\epsilon$ and $b^\epsilon$ such that for $l=1, \ldots, n$
\bea
a^\epsilon_l &= & a^*_l + \epsilon \ a^*_l  \ 1_{l  \in B^1_{i_j} }   \\
b^\epsilon_l  & = & b^*_l
\eea
Let $r \le s \in \{1, \ldots,n \}$. If $r \notin B^1_{i_j}$ and $s \notin B^1_{i_j}$, then $ a^\epsilon_s - a^\epsilon_r = a^*_s - a^*_r \ge 0 $. If $r \in B^1_{i_j}$ and $s \notin B^1_{i_j}$, then $ a^*_s > a^*_r  $ and  $a^\epsilon_s - a^\epsilon_r = a^*_s -a^*_r + \epsilon a^*_s > 0$ for $\vert \epsilon \vert $ small enough. The same reasoning applies if $r \notin B^1_{i_j}$ and $s \in B^1_{i_j}$. Finally, if $r, s \in B^1_{i_j}$, then $a^\epsilon_s - a^\epsilon_r = 0$.

Now, for $r \in \{1, \ldots, n\}$, we have $a^\epsilon_r = a^*_r \le b^*_r $ if $r \notin B^1_{i_j}$. Otherwise, $a^\epsilon_r = a^*_r (1+ \epsilon) < b^*_r$ if $\vert \epsilon \vert $ is small enough. Hence, $(a^\epsilon, b^\epsilon) \in \II_n$, and
\bea
0 &= &\lim_{\epsilon \searrow 0} \frac{1}{\epsilon} (L_2(a^\epsilon,b^\epsilon) - L_2(a^*,b^*)) \\
  & = & \sum_{r=1}^n (a^*_r - y_r) 1_{ r\in B^1_{i_j}} a_r^* w_{1,r}.
\eea
Summing up over all the sets $B^1_{i_j}$ yields the identity in \eqref{Eq2}. We can prove very similarly the identity in \eqref{Eq3}.

Conversely, suppose that $(a^*, b^*) \in \II_n$ satisfies the inequality in \eqref{Ineq2}. For any $(a, b) \in \II_n$, we have
\bea
L_2(a,b) - L_2(a^*,b^*) &= & \frac{1}{2} \sum_{i=1}^n (a_i - a^*_i)^2 w_{1,i} + \frac{1}{2} \sum_{i=1}^n (b_i - b^*_i)^2 w_{2,i} \\
&& + \  \sum_{i=1}^n (a^*_i - y_i)(a_i - a^*_i)w_{1,i} \\
&& +  \  \sum_{i=1}^n (b^*_i - z_i)(b_i - b^*_i)w_{2,i} \\
& \ge & 0.
\eea
We conclude that $(a^*, b^*)$ is the solution of the minimization problem.  \hfill $\Box$

\medskip

\pr{\ of \ Proposition~\ref{Expa1}} Let $\epsilon > 0$ and consider $(a, b) \in \mathbb{R}^n \times \mathbb{R}^n$ such that
\bea
a_i &= & a^*_i - \epsilon \ 1_{i \in \{1,\ldots, t\}}, \  t  \in \{1, \ldots, n \} \\
b_i & = & b^*_i
\eea
for $i= 1, \ldots, n$. For small $\epsilon$, $(a, b) \in \II_n$. Using the characterization in Theorem~\ref{charac2}, it follows that
\bea
\sum_{j=1}^t (a^*_j - y_j) w_{1,j} \le 0
\eea
implying that
\bea
\sum_{j=1}^t (a^*_1 - y_j) w_{1,j} \le 0, \ \ \textrm{for  \ $ t \in \{1,\ldots,n \}$}
\eea
or equivalently
\bea
a^*_1  \le \min_{ t \ge 1 } Av_1(\{1,\ldots, t\}).
\eea
Now, consider $(a, b) \in \mathbb{R}^n \times \mathbb{R}^n$ such that
\bea
a_j &= & a^*_j - \epsilon 1_{j \in \{1,\ldots, t\}}, \   t \in \{1, \ldots, n\}  \\
b_j & = & b^*_j - \epsilon 1_{j \in \{1, \ldots, t'\}}, \  \ 1 \le t' \le t
\eea
for $j=1,\ldots,n$, with $\epsilon > 0$. For small $\epsilon$, we have that $(a, b) \in \II_2$, and hence
\bea
\sum_{j=1}^t (a^*_j - y_j) w_{1,j} +  \sum_{j=1}^t (b^*_j - z_j) w_{2,j} \le 0.
\eea
It follows that
\bea
\sum_{j=1}^t (a^*_1 - y_j) w_{1,j} +  \sum_{j=1}^{t'} (a^*_1 - z_j) w_{2,j} \ge 0,
\eea
that is
\bea
a^*_1 \le \min_{1 \le t' \le t \le  n } \tilde{M}(\{1,\ldots, t\}, \{1,\ldots, t'\}).
\eea
We conclude that
\bea
a^*_1 \le \min_{t \ge 1} Av_1(\{1,\ldots, t\}) \wedge \min_{t \ge t' \ge  1 } \tilde{M}(\{1,\ldots, t\}, \{1,\ldots, t'\}) .
\eea
Now if  $a^*_1 < b^*_1$, let $i_1 \{1, \ldots, n\}$ be such that $a^*_1 = \ldots = a^*_{i_1} $. Then $(a, b)$ is such that
\bea
a_j &= & a^*_j + \epsilon \ 1_{j \in \{1,\ldots, i_1\}} \\
b_j & = & b^*_j
\eea
for $j =1, \ldots, n$ is in $\II_n$ when $\vert \epsilon \vert $ is small enough. It follows that
\bea
Av_1(\{1,\ldots, i_1\}) &=& a^*_1.
\eea
If  $a^*_1 = b^*_1$, and $i'_1$ and $i''_1$ are such that $a^*_1=\ldots=a^*_{i'_1}$ and $b^*_1=\ldots=b^*_{i''_1}$, then $(a, b)$ such that
\bea
a_j &= & a^*_j + \epsilon \ 1_{j \in \{1,\ldots, i'_1 \}} \\
b_j & = & b^*_j +  \epsilon \ 1_{j \in \{1,\ldots,i''_1 \}}
\eea
for $j=1, \ldots, n$ is in $\II_n$ for $\vert \epsilon \vert $ small enough. Hence,
\bea
 a^*_1 = \tilde{M}(\{1, \ldots, i'_1\}, \{1,\ldots, i''_1\}).
\eea
(note that $i''_1 \le i'_1$).  Therefore,
\bea
a^*_1 = \min_{t \ge 1} Av_1(\{1,\ldots, t\}) \wedge \max_{t \ge t' \ge  1 } \tilde{M}(\{1,\ldots, t\}, \{1,\ldots, t'\}).
\eea
The expression of $b^*_1$ follows easily by replacing respectively $y_i$ and $z_i$ by  $-z_{n-i+1}$ and $ -y_{n-i+1}$ for $i=1, \ldots, n$. $\Box$


\medskip

\pr{\ of \ Theorem \ref{PAVA2}} Consider $a \in \mathbb R^n$ given by
\bea
a_i& =& \max_{s \le i} \min_{t \ge i} M(\{s,\ldots, t\})
\eea
and also the subdivision into subsets $S_j=\{i_{j-1}+1,\ldots, i_j\}$ obtained by the PAVA. Let us denote by $G^-$ (resp. $G^+$) the
grid set of indices which correspond to points at the beginning (resp. end) of those subsets; i.e.  of the form $i_{j}+1$ (resp. $i_{j}$).

We obviously have
$$a_i \leq \max_{s\leq i} \min_{t\geq i,\, t\in G^+} M(\{s,\ldots, t\}).$$
Then, consider $s\notin G^-$. This means that we have a set $\{s,\ldots, t\}$ of the form $B\cup C$, $C$  being a union of subsets in the subdivision and $B$ a right subset of a set of the partition
of the form $A \cup B$. We want to prove that $M(\{s,\ldots, t\})=M(B\cup C)$ is either smaller than $M(C)$ or $M(A\cup B\cup C)$. Suppose this is not the case. Then we would have
$$M(B\cup C)>M(C),\;M(B\cup C)>M(A\cup B\cup C),\; M(A)>M(B),$$
where the last inequality is implied by  the second property in Theorem \ref{PAVA}.
Yet, the second inequality, together with the \ap, implies that $M(A)<M(B\cup C)$. In the end we get
$$M(B\cup C)>M(C),\;M(B\cup C)>M(A)>M(B),$$
which contradicts the \ap.

We conclude that  $M(\{s,\ldots, t\})$ is smaller than the value of $M$ at a set which is a union of sets of the subdivision; i.e. either $A\cup B\cup C$ or $C$ itself. But on sets of this kind it is obvious, by the \ap, that $M$ is smaller than the value $m_t$, since this is the maximal value of $M$ on the intervals composing such a set (this is a consequence of $M$ being increasing). Hence, $M(\{x_s,\ldots, x_t\})\leq m_t$, implying that
$$a_i \leq \max_{s\leq i} \min_{t\geq i,\, t\in G^+} m_t =m_i.$$

The opposite inequality is obtained exactly in a symmetric way (first take $s\in G^-$, then prove that $M(\{x_s,\ldots, x_t\})$ is larger than the value of $M$ on a union of intervals).   \hfill $\Box$

\section{Computing the subgradient} \label{subgradient}

\paragraph*{Computing the subgradient of $\Psi$ on a dense set.} Consider the set
\bea
D  &= & \Big  \{b=(b_1, \ldots, b_{n-1}) \in \mathbb{R}^{n-1}: b_i \ne b_j \ \forall \ i \ne j, \  \\
&& \hspace{3.2cm} \textrm{and} \ b_{i'} \ne G_{s,j'} \ \forall \ 1\le i'\le n-1, 1 \le s \le n-1,  1 \le j' \le n \Big \}.
\eea
We denote by $(e^1, \ldots,e^{n-1})$ the canonical basis of $\mathbb{R}^{n-1}$. The set $D$ is a dense open subset of $\mathbb{R}^{n-1}$ where the function $\Psi$ is differentiable. Actually, for a fixed $b \in D$, in the explicit formula for $\Psi$ there is no ex-aequo (up to possible equalities between the $G_{i,s}$ terms). The same will be true in a neighborhood of $b$. For each value of $i \in \{1, \ldots, n \}$, we define the function
\bea
\Psi_i = \Big(\max_{s \le i} (G_{s,i} \wedge b_s) - y_i\Big)^2 w_{1,i}.
\eea
Let us first consider $i \in \{1, \ldots, n-1\}$. We define  $\{s_{i_1}, \ldots, s_{i_k}\}$ to be the set of indices $s$ where $\max_{s \le i} (G_{s,i} \wedge b_s)$ is attained.

\medskip

If $k=1$, then $G_{s_{i_1},i} \wedge b_{s_1} > G_{s,i} \wedge b_s$ for all $ s \in \{1, \ldots, i\} \setminus \{ s_{i_1} \} $.
This implies that the same strict inequalities will be true in a neighborhood of $b$ and hence there are two cases: either the function is locally constant or the square of an affine function. Hence,

\bi
\item If $ b_{s_{i_1}} > G_{s_{i_1}, i}$, then $\nabla \Psi_i(b) = 0$.
\item If  $ b_{s_{i_1}} < G_{s_{i_1}, i}$, then $\nabla \Psi_i(b) = 2\Bl( (G_{s_{i_1},i} \wedge b_{s_{i_1}}) - y_i\Br) \ w_{1,i} \ e^{s_{i_1}}$.
\ei

Now if $k \ge 2$, then this implies that only $G_{s_{i_j}, i}, j=1, \ldots, k$ can be equal (by definition of the set $D$), and hence the function is locally constant.  Therefore, $\nabla \Psi_i(b) = 0$.

For $i=n$, the calculation also requires distinction between the cases $k=1$ and $k \ge 2$. Thus, if $k =1$ and the maximum $\max_{s \le n} (G_{s,n} \wedge b_s)$ is attained at $s_{i_1} \ne n$, then

\bi
\item  If  $ b_{s_{i_1}} > G_{s_{i_1}, n}$, then $\nabla \Psi_i(b) = 0$.
\item If  $ b_{s_{i_1}} < G_{s_{i_1}, n}$, then $\nabla \Psi_n(b) = 2 \Bl((G_{s_{i_1},n} \wedge b_{s_{i_1}}) - y_n \Br) \ w_{1,n} \ e^{s_{i_1}}$.
\ei

If $k=1$ and $s_{i_1} = n$ (in this case $b_n = b^*_n$ is known) or $k \ge 2$, then $\nabla \Psi_n(b) = 0$.
Now the gradient $\nabla \Psi(b)$ is given by
\bea
\nabla \Psi(b) = \sum_{i=1}^n \nabla \Psi_i(b) + 2 \sum_{i=1}^{n -1}(b_i - z_i)w_{2,i} e^i.
\eea

\paragraph*{Calculating the subgradient of $\Psi$ at any point.}

Take now any point $b \in \mathbb{R}^{n-1}$ which does not necessarily belong to $D$. We want to approximate $b$ by points of $D$ in the perspective of using the following property: If $\Psi$ is convex, $p_\varepsilon\to p$, $\gamma_\varepsilon\to \gamma$ as $\epsilon \to 0$,  and $\gamma_\varepsilon\in \partial\Psi(p_\varepsilon)$, then $\gamma\in\partial\Psi(p)$. This is useful when we only want to find one element of the subdifferential at a given point and we already know the gradients at nearby points.

 We use the following approximation:
 $$b_\varepsilon=b+\varepsilon u, \;\mbox{ where } u=(1,2,\dots, i, \dots, n-1).$$
 We claim that $b_\varepsilon$ may belong to the complement of $D$ for a finite number of values $\varepsilon$ at most. Indeed,  for any pair $(i,j)$ with $i\neq j$, the equality $b_i+i\varepsilon=b_j+j\varepsilon$ is satisfied for a unique value of $\varepsilon$, and for any $i,i'$ and $s$, the same thing holds true for the equality $G_{i,s}=b_{i'}+ \varepsilon i'$. Hence, there exists $\varepsilon_0 > 0$ such that for $\varepsilon\in ]0,\varepsilon_0[$, we have $b_\varepsilon\in D$, where the expression of the gradient is fully known by our calculations above.

We can act as follows: Take $b$ and fix $i \le n-1$. For any $s\leq i$, determine which one is minimal among $G_{i,s}$ and $b_s$. In case of equality, priority will be given to $G_{i,s}$ since  in the approximation with $b_\varepsilon$, the value of  $G_{i,s}$ would be smaller than $b_s + \epsilon s$. This way we classify the indices in two categories: The G-type and b-type.  Next, look at all the indices $s_1,\dots,s_k$ realizing the minimum of $G_{i,s}\vee b_s$. If among $s_1, \ldots, s_k$ there are some which are of the b-type, this would imply that in the approximation with $b_\varepsilon$, those indices will yield even a higher value for $G_{i,s_j} \vee (b_{s_j} + \varepsilon s_j)$. In particular the maximal one will correspond to the largest b-type index since it is the one where the coordinate is increased the most in the approximation. Due to the fact that $b^*_n$ is fixed,  we adopt, for $i=n$, the convention that the index $s= n$ is of the G-type when $G_{n,n} \wedge b^*_n$ is maximal. Thus, we can define the vector
 \bea
\tilde{\nabla} \Psi_i(b) &=& 2 ((G_{s_{i_m},i} \wedge b_{s_{i_m}}) - y_i) \ w_{1,i}\ e^{s_{i_m}} \mbox{ or } 0,
\eea
where the index $s_{i_m}$ is the largest index of b-type such that $G_{i,s}\wedge b_s$ is maximal (note that $s_{i_m}$ is always $\le n-1$). If no such index exists (i.e. if the maximal ones are all of G-type), then this is the case where the vector equals $0$. Now consider
\bea
\tilde{\nabla} \Psi(b) &=& \sum_{i=1}^n \tilde{\nabla} \Psi_i(b) + 2 \sum_{i=1}^{n -1}(b_i - z_i)\ w_{2,i} \ e^i.
\eea
This vector belongs to $\partial\Psi(b)$ by approximation and closedness of the subdifferential.

Note that we would have obtained another element of the subdifferential if we had fixed a different order of priority on the coordinates of $b$; for instance the first index instead of the last one (if $u=(1,2,\dots, i, \dots n-1)$ was replaced with $(n-1,\dots, 2, 1))$.  We could also have decreased (instead of increased)  the components, thus giving priority to $b_s$ instead of $G_{i,s}$ in the minimum $G_{i,s}\wedge b_s$. In that case, we would have obtained $0$ for the subgradient of $\Psi_i$ as soon as one of the components realizing the maximum was of the G-type.

\end{appendix}

\end{document}